\documentclass{article}

\usepackage[utf8]{inputenc}
\usepackage[T1]{fontenc}
\usepackage{microtype}
\usepackage{graphicx}
\usepackage{subfigure}
\usepackage{booktabs}
\usepackage{url}
\usepackage{nicefrac}
\usepackage{amsfonts}
\usepackage{float}
\usepackage{enumitem}

\usepackage{hyperref}

\usepackage[accepted]{icml2026}

\usepackage{amsmath}
\usepackage{amssymb}
\usepackage{mathtools}
\usepackage{amsthm}

\usepackage[capitalize,noabbrev]{cleveref}

\theoremstyle{plain}

\theoremstyle{definition}

\theoremstyle{remark}

\icmltitlerunning{Fragility of Explanations in Audio Models}

\begin{document}

\twocolumn[
\icmltitle{The Perceived Fragility of Explanations in Audio Models: Manipulation of Attribution with Unchanged Predictions}

\begin{icmlauthorlist}
\icmlauthor{Piotr Kitłowski}{pw}
\icmlauthor{Dominik Wiącek}{pw}
\icmlauthor{Mateusz Modrzejewski}{pw}
\end{icmlauthorlist}

\icmlaffiliation{pw}{Faculty of Electronics and Information Technology, Warsaw University of Technology, Warsaw, Poland}

\icmlcorrespondingauthor{Piotr Kitłowski}{01149528@pw.edu.pl}
\icmlcorrespondingauthor{Dominik Wiącek}{01169141@pw.edu.pl}
\icmlcorrespondingauthor{Mateusz Modrzejewski}{mateusz.modrzejewski@pw.edu.pl}

\icmlkeywords{Explainable AI, Audio Classification, Attribution Maps, Adversarial Perturbations, Perceptual Metrics}

\vskip 0.3in
]

\printAffiliationsAndNotice{}

\begin{abstract}
This paper investigates the fragility of post-hoc explanation methods in audio deepfake detection. While previous work on explanation manipulation focused on images using standard $L_p$ metrics, we introduce a psychoacoustic framework that optimizes inaudible perturbations to decouple model attributions from final classifications. We evaluate this vulnerability across state-of-the-art architectures under strict prediction-preserving constraints. By evaluating the manipulation cost through domain-specific perceptual audio quality metrics alongside explanation alignment criteria, our framework demonstrates that an adversary can systematically distort automated explanation heatmaps while preserving the predicted deepfake label. Full code available at: https://github.com/cncPomper/Audio-XAI
\end{abstract}

\section{Introduction}
The proliferation of synthetic audio, driven by advances in generative models, has made deepfake detection critical. To foster trust in these systems, Explainable AI (XAI) methods are deployed to highlight the acoustic artifacts driving a model's decision. However, this reliance introduces a security vulnerability: the fragility of the explanations themselves. If an adversary can manipulate the system to provide a deceptive justification while maintaining a prediction, the credibility of the interpretability framework is compromised.

While the fragility of attribution maps has been extensively demonstrated in the computer vision domain \citep{ghorbani2019interpretation, dombrowski2019explanations, heo2019foolingneuralnetworkinterpretations}, its implications for audio models remain largely unaddressed. Furthermore, vision-based attacks measure the manipulation cost using $L_p$ norms, which do not correlate with human auditory perception \citep{abdullah2021beyond}. In the audio domain, an attack on interpretability is viable if the adversarial perturbation remains imperceptible to the ear. Early investigations into the plausibility of audio explanations under adversarial conditions \citep{prinz2023constructing} highlighted the need for more domain-specific, perceptually bounded constraints. More recently, efforts to establish robust baselines for audio deepfake explanations have exposed structural limitations in post-hoc methods like LRP \citep{grinberg2025data}. We expand on this line of thought by demonstrating that these limitations extend beyond inherent inaccuracy; rather, the explanations themselves can be systematically and imperceptibly manipulated by an adversary.

To bridge this gap, this study investigates the stability of post-hoc explanation methods against masked perturbations. We aim to determine whether XAI techniques provide robust interpretations of audio data or can be decoupled from the classifier's decision boundary.

Our main contributions are as follows:
\begin{itemize}[noitemsep, topsep=2pt, parsep=0pt, partopsep=0pt, leftmargin=*]
    \item We introduce a novel optimization framework that successfully adapts explanation-targeted adversarial attacks to the audio domain.
    \item We incorporate a dynamic psychoacoustic masking threshold into the loss function, encouraging large attribution-map changes while penalizing perturbations that exceed a psychoacoustic masking threshold or change the model's final prediction.
    \item We provide a focused empirical evaluation across diverse architectures using the SONICS deepfake dataset \cite{rahman2024sonics}, leveraging domain-specific perceptual metrics to assess perceptual transparency.
\end{itemize}

\section{Methodology}

\subsection{Analyzed XAI Methods}

We investigate two paradigms of post-hoc attribution: \textbf{Grad-CAM} \citep{selvaraju2017grad} and \textbf{Layer-wise Relevance Propagation (LRP)} \citep{bach2015pixel}. While Grad-CAM highlights regions using final-layer gradients, LRP follows a conservation rule to backpropagate relevance scores to the input spectrogram. Relying on both first-order (Grad-CAM) and structurally constrained (LRP) methods allows us to examine whether explanation fragility appears across audio architectures rather than being an artifact of a single algorithm.

\subsection{Dataset and Target Models}
All experiments were conducted using the \textbf{SONICS} dataset, a large-scale corpus designed for audio deepfake detection. To ensure the observed explanation fragility is not an artifact of a single architecture, we evaluated three models employing distinct feature extraction paradigms on time-frequency representations. As a baseline for classical convolutional architectures that rely on local spectrogram patterns, we used \textbf{VGGish} \citep{hershey2017cnn}. To represent modern, self-attention-based architectures capable of modeling global context, we selected the \textbf{Audio Spectrogram Transformer (AST)} \citep{gong2021ast}. Finally, we used \textbf{SpecTTTra}, denoted as \texttt{spectttra-gamma-5s} \citep{rahman2024sonics}, a recent attention-based architecture specifically proposed alongside the SONICS dataset for detecting synthetic music by capturing long-range temporal dependencies. Analyzing this diverse set enables a comparative assessment of the stability of attribution maps across convolutional and token-based processing regimes.

\subsection{Audio Sample Perturbation}


We randomly sampled 100 recordings from the dataset, independent of the label, and applied three attack methodologies targeting the input space $x$ with a bounded perturbation $\delta$. Since attacking attribution mechanisms requires second-order derivatives, the Adam optimizer was used across all methods to ensure convergence.

\textbf{1. Standard Projected Gradient Descent (PGD):} As a baseline, we implemented an $L_{\infty}$-bounded PGD attack \citep{madry2018towards} aimed solely at minimizing the structural similarity between the original and perturbed attribution maps, without considering perceptual audio quality.

\textbf{2. X-Shift Attack (Adapted):} Originally designed for vision-language models \cite{babadi2026xshift}, we adapted this spatial displacement strategy to the audio domain. The objective forces the explanation to assign maximum relevance to a designated, irrelevant target region $M_{target}$, steering it away from the originally salient time-frequency patches $M_{orig}$. 

\textbf{3. Psychoacoustic Noise Modeling (Ours):} To constrain perturbations according to a psychoacoustic masking model, we propose a custom optimization framework. The total loss function $\mathcal{L}(\delta)$ forces map displacement while maintaining rigorous acoustic and predictive constraints:

\begin{equation}
\begin{aligned}
\mathcal{L}(\delta) = \,&\mathcal{L}_{explain}(\delta) \\
&+ \lambda_{aud}\mathcal{L}_{audibility}(\delta) \\
&+ \lambda_{pred}\mathcal{L}_{pred\_preserve}(\delta)
\end{aligned}
\end{equation}

The $\mathcal{L}_{explain}(\delta)$ term minimizes the cosine similarity between the original and perturbed attribution maps. Crucially, to operationalize a psychoacoustic audibility constraint, the threshold penalty is formalized as $\mathcal{L}_{audibility}(\delta) = \mathbb{E}[\max(0, 20\log_{10}|\mathcal{F}(\delta)| - T(x))^2]$, where $T(x)$ is the static masking threshold pre-computed from the clean input. This exclusively penalizes the spectral energy of $\delta$ that exceeds the bounds of human perception. Because manipulating explanations requires differentiating through gradients (second-order derivatives), we optimize the total loss using Adam rather than standard sign-based methods. This is coupled with a margin-based hinge loss ($\mathcal{L}_{pred\_preserve}$) to penalize changes in the original prediction and a hard waveform amplitude constraint $\delta \in [-\varepsilon, \varepsilon]$.

\subsection{Evaluation Metrics}
To assess perceptual transparency, audio fidelity was evaluated across the SONICS dataset using \textbf{PEAQ} \citep{thiede2000peaq}, \textbf{ViSQOL} \citep{chinen2020visqol}, \textbf{Zimtohrli} \cite{alakuijala2025zimtohrli}, \textbf{CDPAM} \citep{manocha2021cdpam}, \textbf{PESQ} \citep{rix2001perceptual}, and \textbf{STOI} \citep{taal2011algorithm}. High values across these indicators corroborate the minimal energy of the injected noise. Finally, explanation fragility was quantified by measuring the discrepancy between original and perturbed attribution maps using \textbf{Cosine Similarity} and \textbf{Top-10 Overlap}.
\subsubsection{Audio Fragility Score}
To summarize the vulnerability of explanation maps, we introduce the Audio Fragility Score (\textit{AFS\textsubscript{stable}}). Unlike binary attack success rates, \textit{AFS\textsubscript{stable}} offers a continuous measure of attribution displacement, conditioned on preserving the predicted class and maintaining high perceptual quality. For a given sample $i$, the metric is defined as:
\vspace{-2mm}
\begin{equation}
\begin{aligned}
AFS^{stable}_i
&= \left(1 - \frac{C_i + T_i}{2}\right)
   \mathbf{1}[\hat{y}^{orig}_i = \hat{y}^{adv}_i] Q_i, \\
C_i &= \cos(A^{orig}_i, A^{adv}_i), \\
T_i &= \mathrm{Top10}(A^{orig}_i, A^{adv}_i).
\end{aligned}
\end{equation}
The first term measures the magnitude of the explanation shift by averaging the cosine similarity and Top-10 overlap between the original ($A^{orig}_i$) and perturbed ($A^{adv}_i$) attribution maps. The indicator function $\mathbf{1}[\cdot]$ acts as a strict penalty, zeroing the score if the predicted class $\hat{y}$ changes. Finally, $Q_i \in [0, 1]$ represents the normalized perceptual quality score of the adversarial audio. Consequently, an $AFS^{stable}$ approaching $1.0$ signifies a highly successful, imperceptible manipulation of the attribution map. In contrast, a score of $0.0$ indicates a failure to shift the explanation, an altered model prediction, or unacceptable acoustic degradation.
\vspace{-4mm}
\section{Results}
\subsection{Global Results}
A practical attack on explainability must remain imperceptible. As shown in Table \ref{tab:quality_metrics}, unconstrained optimization methods such as \textbf{PGD} severely degrade audio quality (e.g., PESQ $\approx 2.8$) and introduce easily audible artifacts. While \textbf{X-Shift} maintains acceptable fidelity, our \textbf{Psychoacoustic} framework preserves high objective perceptual quality (ViSQOL $> 4.1$, CDPAM $\ge 0.98$) by bounding noise within human masking thresholds. These results suggest that explanations for deepfake detection can be substantially manipulated while preserving predictions and maintaining high objective perceptual quality.
\begin{table}[htbp]
\centering
\caption{Median perceptual quality metrics for adversarial audio samples across evaluated models and attack strategies.}
\label{tab:quality_metrics}
\resizebox{\columnwidth}{!}{%
\begin{tabular}{llcccccc}
\toprule
\textbf{Model} & \textbf{Attack} & \textbf{PESQ $\uparrow$} & \textbf{STOI $\uparrow$} & \textbf{ViSQOL $\uparrow$} & \textbf{PEAQ $\uparrow$} & \textbf{Zimtohrli $\uparrow$} & \textbf{CDPAM $\uparrow$} \\
& & \textbf{[-0.5, 4.5]} & \textbf{[0, 1]} & \textbf{[1, 5]} & \textbf{[-4, 0]} & \textbf{[0, 5]} & \textbf{[0, 1]} \\
\midrule
AST       & Psychoacoustic (Ours) & 4.06 & 0.987 & 4.64 & -2.00 & 4.42 & 0.989 \\
          & PGD            & 2.77 & 0.952 & 3.80 & -3.39 & 3.10 & 0.858 \\
          & X-Shift        & 3.87 & 0.990 & 4.46 & -1.80 & 3.47 & 0.950 \\
\midrule
SpecTTTra & Psychoacoustic (Ours) & 3.77 & 0.960 & 4.15 & -2.55 & 3.79 & 0.981 \\
          & PGD            & 2.76 & 0.950 & 3.79 & -3.39 & 3.14 & 0.859 \\
          & X-Shift        & 3.74 & 0.993 & 4.48 & -2.10 & 3.70 & 0.925 \\
\midrule
VGGish    & Psychoacoustic (Ours) & 4.43 & 0.997 & 4.89 & -0.41 & 4.62 & 0.995 \\
          & PGD            & 2.84 & 0.953 & 3.86 & -3.37 & 3.22 & 0.842 \\
          & X-Shift        & 3.78 & 0.990 & 4.31 & -2.16 & 3.56 & 0.938 \\
\bottomrule
\end{tabular}%
}
\end{table}
To systematically compare the vulnerability of the models and the efficacy of the adversarial strategies, we aggregated the sample-level $AFS^{stable}$ scores into a unified ranking system. For each test sample, the model-attack combinations were ranked based on their explanation displacement, where lower ranks indicate higher vulnerability. As detailed in Table \ref{tab:rankings}, \textbf{SpecTTTra} demonstrated the highest resistance to explanation manipulation (Mean Rank: $7.83 \pm 0.48$). Its mechanism of tracking long-range temporal dependencies seemingly dilutes the impact of constrained adversarial noise. Conversely, the token-based \textbf{AST} was the most fragile architecture ($3.00 \pm 0.58$), allowing adversaries to easily manipulate its attention maps.
\begin{table}[htbp]
\centering
\caption{Overall robustness rankings across evaluated architectures and attack strategies. A lower rank indicates higher vulnerability (i.e., easier to manipulate the explanation).}
\label{tab:rankings}
\begin{tabular}{lcc}
\toprule
\textbf{Configuration} & \textbf{Median Rank} & \textbf{Mean Rank ($\pm$ SD)} \\
\midrule
SpecTTTra      & 8.0 & $7.83 \pm 0.48$ \\
VGGish         & 4.5 & $4.17 \pm 0.95$ \\
AST            & 3.0 & $3.00 \pm 0.58$ \\
\midrule
X-Shift        & 6.0 & $5.83 \pm 1.28$ \\
PGD            & 5.5 & $5.00 \pm 0.68$ \\
Psychoacoustic & 3.0 & $4.17 \pm 1.28$ \\
\bottomrule
\end{tabular}
\end{table}

\subsection{Visual Comparison of Sample Maps}
\begin{figure}[t]
    \centering
    \includegraphics[width=\columnwidth]{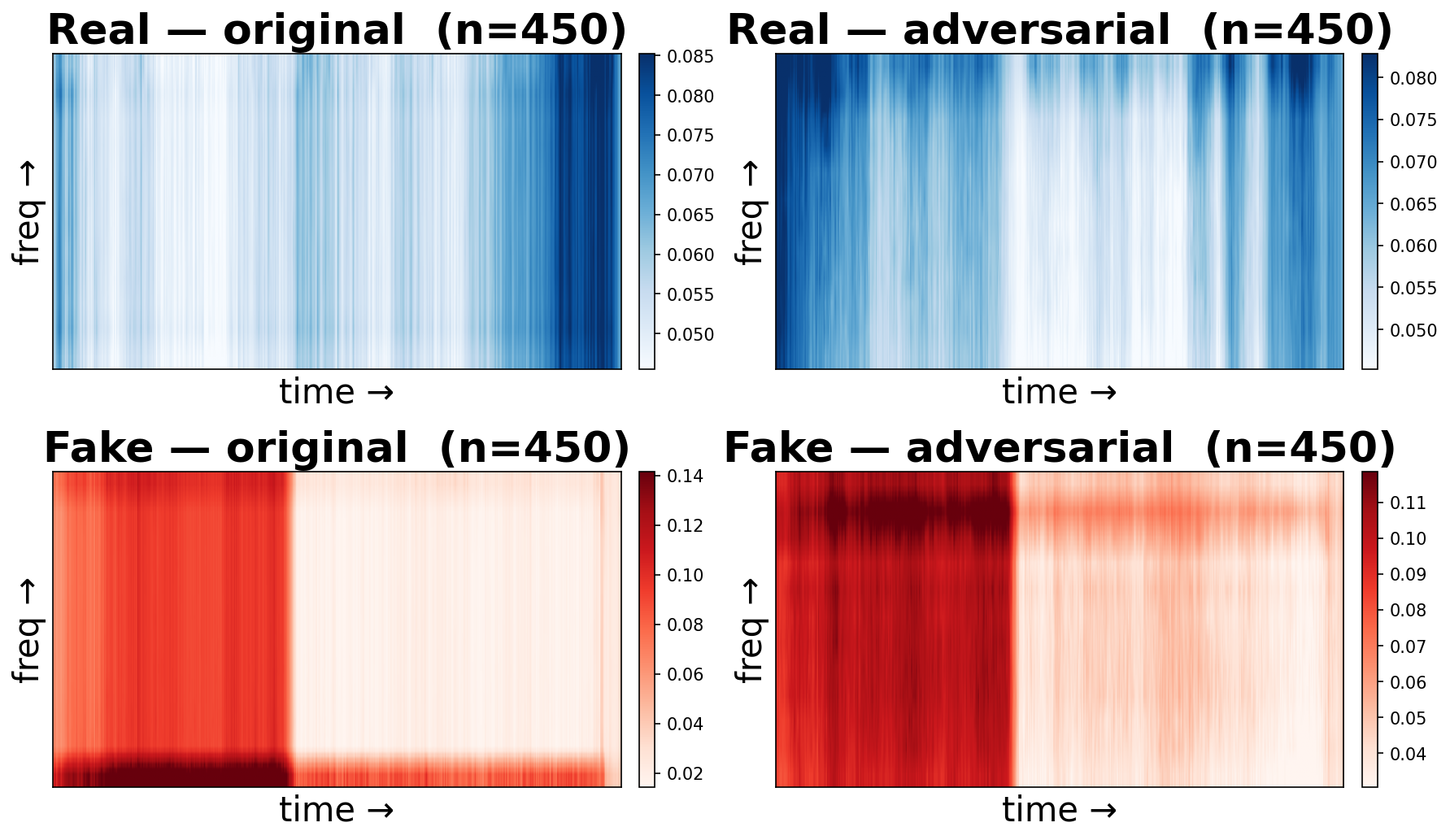}
    \caption{Average attribution heatmaps demonstrate that the adversarial attacks systematically distort and diffuse the original time-frequency explanation structures for both real and fake audio samples while strictly preserving the model's final classifications.  }
    \label{fig:gradcam_rf}
    \end{figure}

\begin{figure}[t]
    \centering
    \includegraphics[width=\columnwidth]{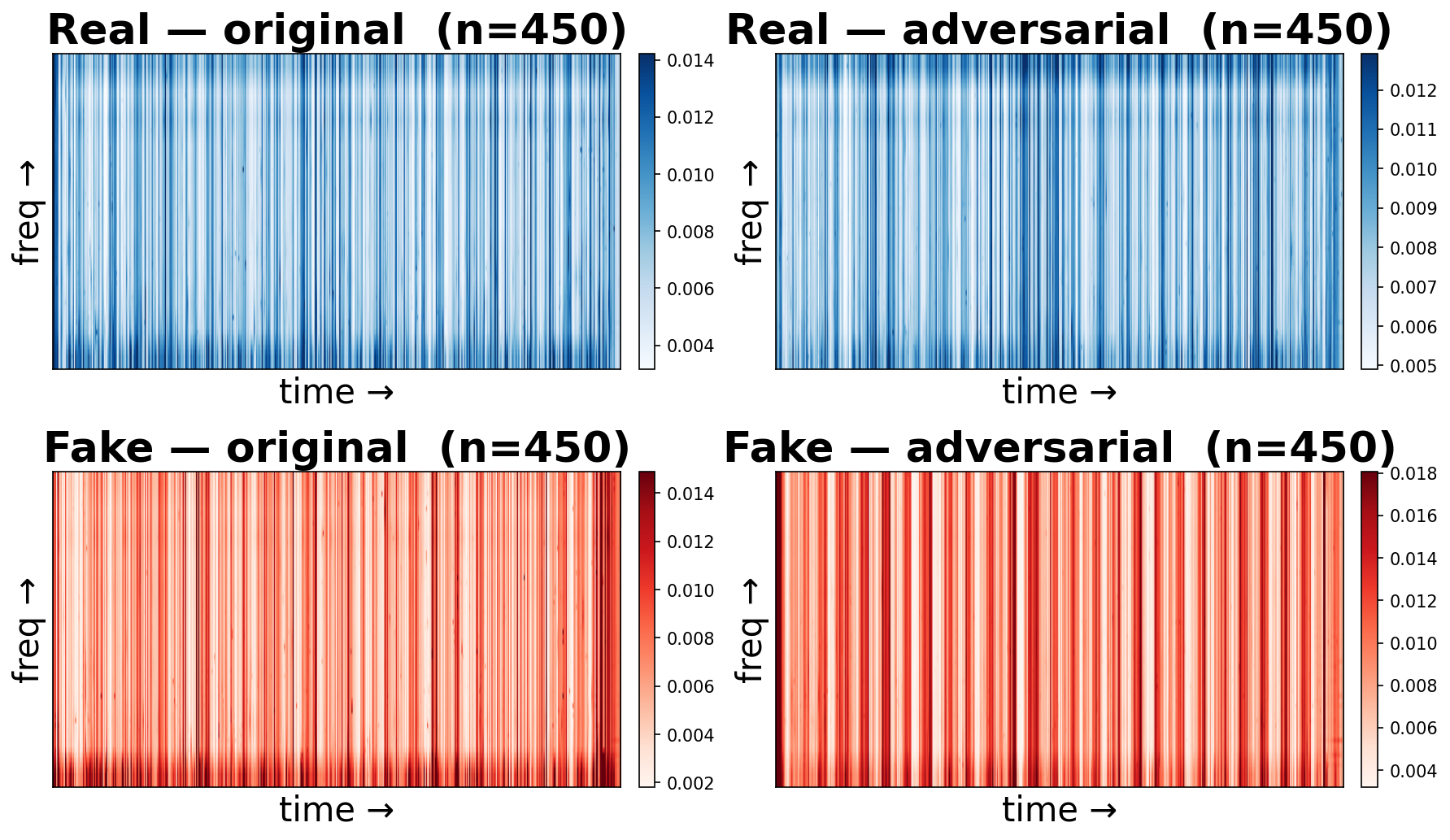}
    \caption{Average Layer-wise Relevance Propagation (LRP) attribution maps show that adversarial attacks systematically diffuse and distort time-frequency explanation structures for both real and fake audio while preserving the model's classifications.}
    \label{fig:lrp_rf}
\end{figure}

    \textbf{Complementary Interpretations:}
    LRP reveals high-resolution in Fig.~\ref{fig:lrp_rf}, frame-by-frame attributions concentrated in acoustic low-frequency signatures, while Grad-CAM in Fig.~\ref{fig:gradcam_rf} shows how these features are aggregated into macro-temporal windows (e.g., asymmetric temporal bias between early and late segments) in deeper layers.
    \textbf{Adversarial Mechanism Insight:}
    Adversarial attacks exploit both abstraction levels: they introduce periodic, pixel-level perturbations across the audio timeline (LRP) that shift the model's global attention windows (Grad-CAM), blinding it to truly discriminative acoustic regions.
\vspace{-3mm}
\subsection{Comparison of individual samples}
\begin{figure}[H]
    \centering
    \includegraphics[width=\columnwidth]{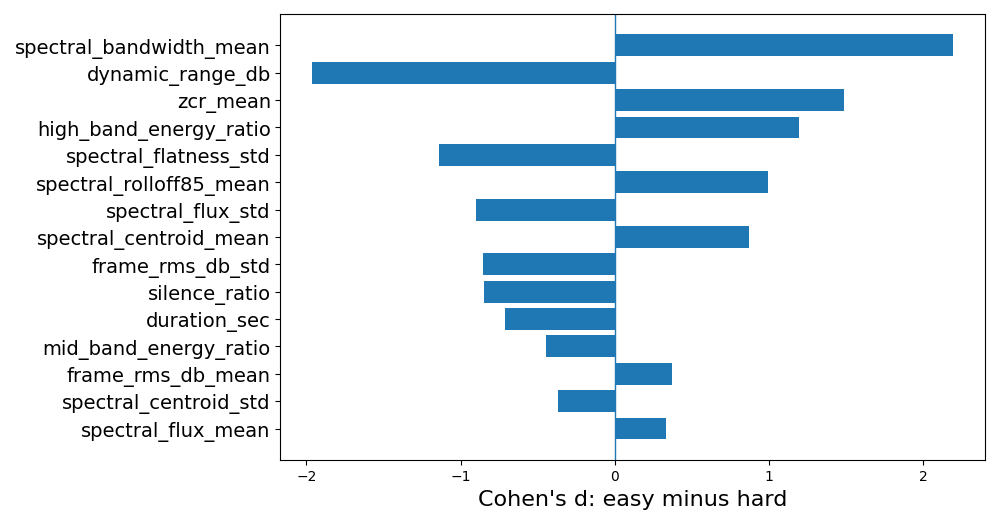}
    \caption{Bar chart of the audio parameters that differ most between ``easy'' and ``hard'' samples}
    \label{fig:ranking_indiv}
\end{figure}

Given our AFS stable metric, we sorted all analyzed samples by this metric, extracting top 10 "easiest" and "hardest" samples to attack, all of them retaining high audio quality and original ML class prediction, where "easy" samples had the greatest changes to map attribution, while "hard" ones had the smallest changes. We then identified audio parameters in which these two groups differed the most, as seen in Fig.~\ref{fig:ranking_indiv}:
\raggedbottom
\vspace{-6mm}
Vulnerability appears associated with dense, broadband audio characteristics: ``easy'' samples exhibit higher spectral bandwidth, zero-crossing rates, and high-frequency energy (rock/electronic music). These ``busy'' textures may yield larger estimated masking budgets, giving optimizers an imperceptible budget to displace attribution maps. Conversely, ``hard'' samples are defined by extreme acoustic sparsity - high dynamic range and frequent silences (classical, acoustic music) - where strict perceptual constraints severely limit allowable adversarial noise.
\vspace{-2mm}
\subsection{PCA evaluation}
\begin{figure}[H]
    \centering
    \includegraphics[width=\columnwidth]{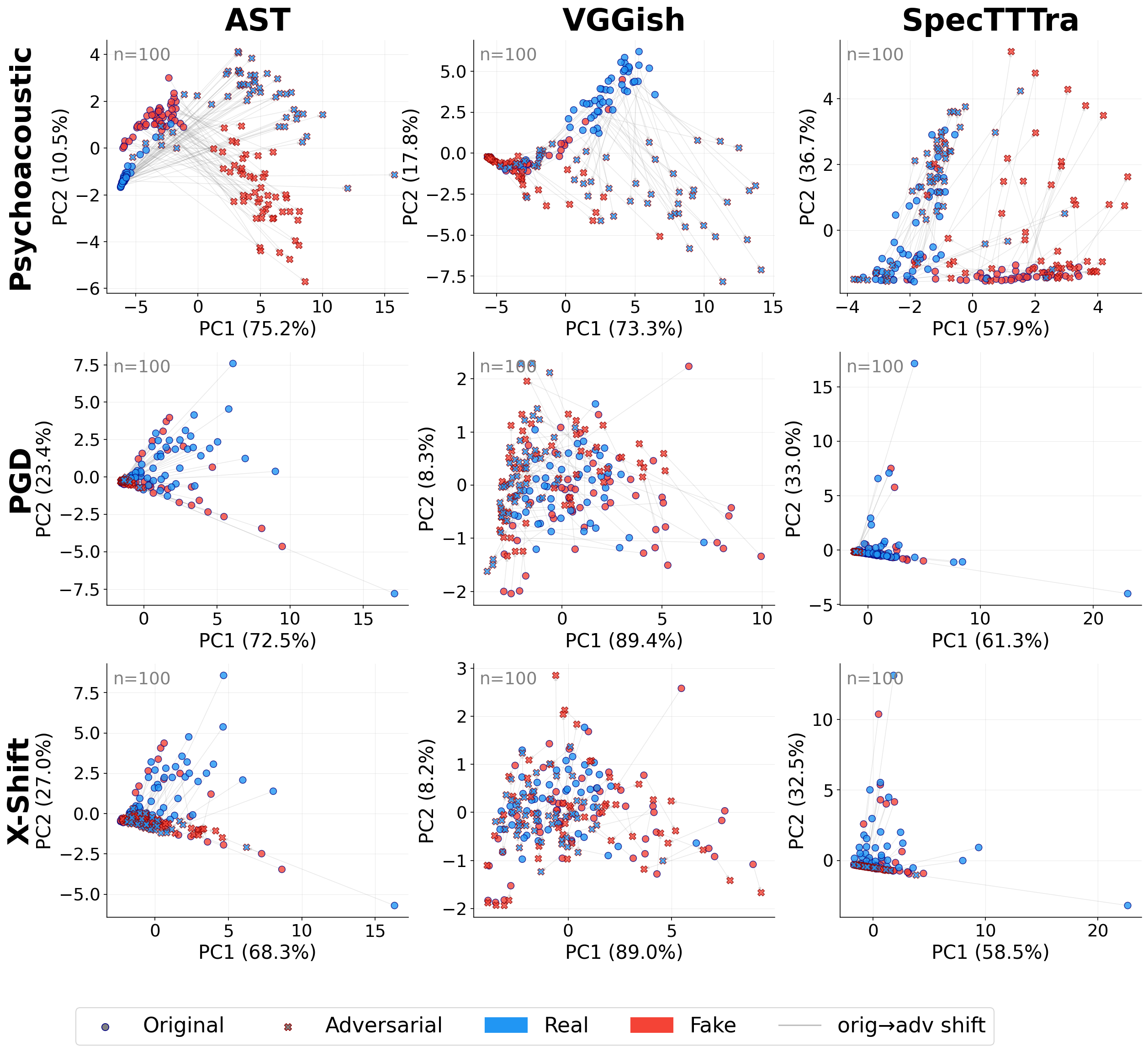}
    \caption{PCA maps for combinations model x attack}
    \label{fig:pca9}
\end{figure}

\vspace{-2mm}
Our geometric analysis in Fig.~\ref{fig:pca9} demonstrates that the psychoacoustic attack forces smooth, directional attribution shifts for transformer-based models, while maintaining structural dispersion in the convolutional model.
Conversely, unconstrained PGD and X-Shift attacks primarily induce variance reduction and cluster compression rather than steered displacement. These distinct patterns confirm that attention-based architectures are highly susceptible to systematic explanation steering, raising concerns about using post-hoc attribution maps as standalone auditing tools for audio deepfake detectors.

\section{Evaluation and Conclusions}


This study provides evidence that post-hoc explanations in audio deepfake detection are fragile under targeted perturbations. Leveraging our psychoacoustic optimization framework, we demonstrate that attribution maps---such as Grad-CAM and LRP \cite{heo2019foolingneuralnetworkinterpretations}---can be systematically manipulated without altering predictions or degrading perceptual audio fidelity. Unlike unconstrained $L_p$ attacks, which introduce audible artifacts, our dynamic masking approach reveals a security gap: explanations can be decoupled from prediction-preserving behavior through perturbations within objective psychoacoustic constraints.


Findings reveal architectural and acoustic dependencies. While recent work exposed structural limitations in post-hoc explanations \citep{grinberg2025data}, we show these weaknesses enable deliberate adversarial manipulation. In PCA analysis, attention-based models exhibited directional shifts in attribution space, proving more vulnerable to steering than convolutional networks. Furthermore, acoustic topology dictates robustness: dense, broadband signals provide a masking budget for noise, whereas sparse tracks with high dynamic range restrict optimization. Relying on visual attributions for audio model trustworthiness is premature; future research must develop interpretability mechanisms mathematically tethered to the classifier's exact decision boundary.

\section*{Impact Statement}
This paper investigates vulnerabilities in explanations produced by audio deepfake detection systems. The work may help improve the robustness and trustworthiness of deployed detection pipelines, but it also studies manipulation mechanisms that could be misused to obscure model behavior. We therefore frame the attacks as a diagnostic tool for developing more reliable explanation methods and recommend that any released code or artifacts include safeguards, clear documentation, and evaluation protocols for defensive use. We release the full code, configurations, attack hyperparameters, preprocessing, and evaluation scripts to support reproducibility and responsible use.

\section{Acknowledgments}
We gratefully acknowledge Polish high-performance computing infrastructure PLGrid (HPC Center: ACK Cyfronet AGH) for providing computer facilities and support within computational grant no. PLG/2026/019417.

\bibliography{references}
\bibliographystyle{icml2026}

\end{document}